\documentclass[twocolumn,prb,aps,showpacs]{revtex4}
\usepackage{graphicx}
\usepackage{amsmath}

\begin{document}
\preprint{in preparation}

\title[Electronic Stopping of Diamond]{Electronic Stopping and Momentum Density of Diamond
Obtained from First-Principles Calculations}

 \author{Richard J. Mathar}
 \email{mathar@mpia-hd.mpg.de}
 \affiliation{Goethestr.\ 22, 69151 Neckargem\"und, Germany }

\date{\today}

\begin{abstract}
We calculate the ``head'' element or the
$({\bf 0,0})$-element of the wave-vector and frequency-dependent
dielectric matrix of bulk crystals via first-principles,
all-electron Kohn-Sham states in the integral of the irreducible polarizability
in the random phase approximation.
We approximate the macroscopic ``head'' element of the inverse matrix by its
reciprocal value, and integrate over frequencies and momenta
to obtain the electronic energy loss of protons at low velocities.
Numerical evaluation for diamond targets predicts that the band gap
causes a strong non-linear reduction of the electronic stopping power
at ion velocities below $0.2$ atomic units.
\end{abstract}
\maketitle

\section{Formulation and Method}

The objective of this research is a quantitative, first-principles
description of the energy deposition by a bare ion in diamond.
The energy loss per unit path length of a massive, punctiform, charged particle
with charge number $Z_1$ to a target is\cite{fulltext}
\begin{equation}
\frac{dE}{dx}({\bf v})=\frac{(Z_1e)^2}{4\pi^3\epsilon_0v}
\int d^3q \frac{{\bf v}\cdot{\bf q}}{q^2}\int_0^{\infty}\!\!\!
d\omega \delta(\omega-{\bf q}\cdot {\bf v})
\mathrm{Im}K_{\bf 0,0}({\bf q},\omega),
\label{dedx}
\end{equation}
with
${\bf v}$ the projectile velocity in the target rest frame,
$e$ the elementary charge unit,
and $\epsilon_0$ the vacuum permittivity.
For reciprocal lattice vectors ${\bf G}$,
$K_{\bf 0,0}$ is the ${\bf G}={\bf G}'={\bf 0}$ (``head'') component of 
the inverse dielectric matrix defined by
\[
\sum_{\bf G'}\epsilon_{\bf G,G'}({\bf q},\omega)
K_{\bf G',G''}({\bf q},\omega)=\delta_{\bf GG''}
\]
with respect to $\epsilon_{{\bf G},{\bf G}'}({\bf k},\omega)$, the
wave-vector and frequency dependent microscopic dielectric matrix.
In terms of the irreducible polarizability $\Pi$,
the dielectric matrix is
\[
\epsilon_{\bf G,G'}({\bf q},\omega)=\delta_{\bf GG'}-\frac{e^2}{\epsilon_0
|{\bf q}+{\bf G}|^2}\Pi({\bf q}+{\bf G},{\bf q}+{\bf G'},\omega),
\]
given in the Random Phase Approximation\cite{Adler} (RPA) as a sum over all
band pairs $(\nu,\nu')$ and an integral over the first Brillouin zone (BZ)
\begin{eqnarray}
&&\Pi({\bf q}+{\bf G},{\bf q}+{\bf G'},\omega) \nonumber \\
&&=\sum_{\nu\nu'}\int_{\text{BZ}}\frac{d^3k}{(2\pi)^3}
\frac{m_{\bf G}^* m_{\bf G'}(f_{\nu{\bf k}}-f_{\nu'{\bf k+q}})}
{\hbar \omega+i\eta+E_{\nu{\bf k}}-E_{\nu'{\bf k+q}}} .
\label{Piint}
\end{eqnarray}
$E_{\nu,{\bf k}}$ are the
band energies, and $f_{\nu,{\bf k}}$ are Fermi occupation numbers
($f_{\nu,{\bf k}} =0$ or 2 for $E_{\nu,{\bf k}}$ above
or below the Fermi energy $E_F$).
The matrix elements in Eq.\ (\ref{Piint}) are
\[
m_{\bf G}\equiv \langle \nu'{\bf k+q}|e^{i({\bf G+q})\cdot{\bf r}}|
\nu {\bf k}
\rangle .
\]

We determine the ground-state electronic structure of solids
within Density Functional Theory (DFT) as established in the
Kohn-Sham (KS) variational procedure
and implemented in the computational package {\sc gtoff}.\cite{GTOFF}
The results of the all-electron, full-potential calculations
are eigenfunctions $\varphi_{\nu,{\bf k}}({\bf r})$,
expressed as linear combination of Gaussian Type Orbitals (GTO's),
and eigenvalues $E_{\nu,{\bf k}}$.

The Bloch functions are finally expanded in a (truncated) plane wave (PW)
series,
\[
\varphi_{\nu,{\bf k}}=
\langle {\bf r}|\nu {\bf k}\rangle
= e^{i{\bf k}\cdot{\bf r}}u_{\nu{\bf k}}({\bf r})
= \frac{1}{V_{\text{UC}}}\sum_{\bf G}
e^{i({\bf k+G})\cdot{\bf r}}u_{\nu{\bf k},{\bf G}},
\]
to represent $m_{\bf G}$ as a simple sum over products of
expansion coefficients $u_{\nu{\bf k},{\bf G}}$,\cite{fulltext}
where $V_{\text{UC}}$ is the volume of a unit cell (UC).
The equivalence of the GTO and PW representations is maintained
by monitoring the accumulated norm for each $|\nu,{\bf k}\rangle$ relative
to the exact values, 
\begin{equation}
\sum_{\bf G}|u_{\nu,{\bf k},{\bf G}}|^2=V_{\text{UC}}.
\label{sumPW}
\end{equation}

We subdivide the integration region of the integral (\ref{Piint}) into
${\bf k}$-space tetrahedra. Recursive further subdivision of a given
tetrahedron into smaller tetrahedra is done if
$f_{\nu{\bf k}}-f_{\nu'{\bf k+q}}$
is not constant over all four vertices.
Next comes
linearization of the product of the matrix elements in the numerator
and of the energy denominator inside each tetrahedron for each $\omega$.
The resulting approximated integral is evaluated analytically.\cite{fulltext}

The product $|{\bf q+G}|^2\epsilon_{\bf G,G'}({\bf
q},\omega)$ is calculated and, in compensation, the term $q^2$ in the
denominator of Eq.\ (\ref{dedx}) is dropped.
The dielectric function is tabulated for ${\bf q}$ commensurate with
the uniform mesh of wave vectors used in the underlying {\sc gtoff}
calculation but covering higher BZ's as well as the first.
[Values outside the $({\bf q},\omega)$-meshes are replaced by the vacuum
response, $\mathrm{Im} K_{\bf 0,0}({\bf q},\omega)=0$.]
$\mathrm{Im} K_{\bf G,G'}$ is linearized inside each ${\bf q}$-space
tetrahedron.
Multiplied by the linear factor ${\bf q}\cdot{\bf v}$, the integrals
(\ref{dedx}) over tetrahedra are done analytically, then summed.

\section{Diamond}

\subsection{Basis Sets, Electron Momentum Density}

Diamond is studied at the experimental lattice parameter,
$a=6.74071a_0$ for the cubic unit cell, where $a_0$ denotes one bohr.
The Moruzzi-Janak-Williams parameterization\cite{Moruzzi}
of the Hedin-Lundqvist local-density approximation (LDA) to the
exchange-correlation potential is used.
Small, highly contracted basis sets as used in Ref.\ \onlinecite{Ayma}
are generally insufficient to calculate the real parts of dielectric functions 
and subsequently the energy loss functions.
We started from Partridge's $16s11p$ set,\cite{PartrC} contracted
the seven tightest $s$-functions and the four tightest $p$-functions,
removed the most diffuse $s$ and the two most diffuse $p$-functions to avoid
approximate linear dependencies, and added a full set of three $d$-functions
with exponents equal to the remaining most diffuse $p$-functions.
(Without $d$-orbitals the total energy would rise by 0.24 eV/atom.)
Site centered $s$- and $f$-type fitting functions were used with exponents
for the $s$-types as in Ref.\ \onlinecite{TrickeyC}, and for
the $f$-types $0.5/a_0^2$ and $0.2/a_0^2$ as in Ref.\ \onlinecite{Lu}.
Space group and lattice type are those of silicon; hence we may
refer to a prior {\sc gtoff} study\cite{BoeSi} for the symmetry properties
of the fitting functions.

\begin{figure}[h]
 
 \includegraphics[scale=0.48]{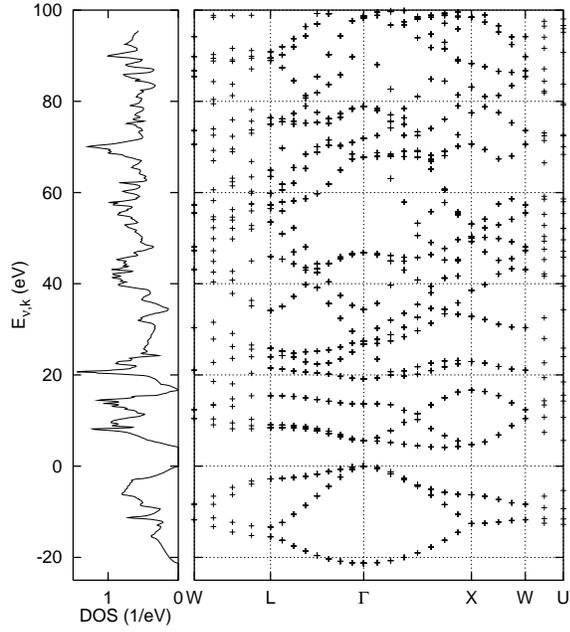}
\caption{
KS band-structure $E_{\nu,{\bf k}}$
using
$16\times 16\times 16$ ${\bf k}$-points in the BZ (213 in the IBZ). The direct
gap at $\Gamma$ is 5.59 eV, the indirect gap 4.13 eV, and the width of the four
valence bands with eight electrons 21.29 eV\@.
The two bands with four core electrons at $-263$ eV are not
shown.\protect\cite{Nithiandam}
}
\label{fig1.ps}
\end{figure}

The density of states (DOS) computed with this $9s6p3d$ basis (84 basis
functions in the primitive UC) as shown in Fig.\ \ref{fig1.ps}
is stable up to $\approx 70$ eV above the Fermi energy
with respect to further de-contraction.
Band gaps are too small compared with experiments, as usual
for the LDA and known from other DFT calculations
on diamond.\cite{diabands}

\begin{figure}[h]
 
 \includegraphics[scale=0.48]{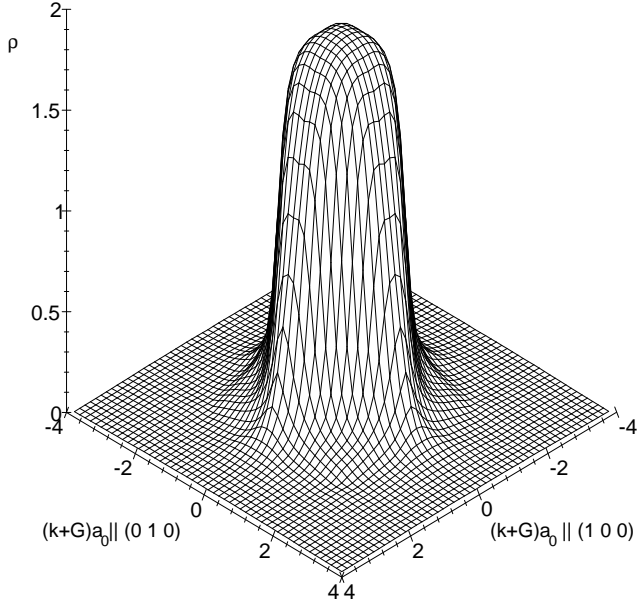}
\caption{
The EMD as a function of ${\bf k+G}$,
which is varied in the plane spanned by $(100)$ and
$(010)$. Both momentum components are measured in units of $1/a_0$.
Anisotropies of ``projected'' positron-EMD's
are discussed in Ref.\ \protect\onlinecite{Nilen}.
}
\label{fig2.ps}
\end{figure}

The lowest 24 bands were expanded into 531 plane waves,
a cut-off at $|{\bf G}|= 7.3/a_0$, with a norm in Eq.\ (\ref{sumPW})
above $0.85V_{\text{UC}}$ for the $K$-shell electrons (excitations
from which were excluded in the subsequent calculations of
$\epsilon_{\bf G,G'}$ and $dE/dx$ anyway),
and a norm above $0.98V_{\text{UC}}$ for the remaining 22 bands.
A first result is the all-electron momentum density (EMD)
\[
\rho({\bf k+G})\equiv \frac{1}{V_{\text{UC}}}\sum_{\nu} f_{\nu,{\bf k}}
|u_{\nu,{\bf k},{\bf G}}|^2
\]
in Fig.\ \ref{fig2.ps}\@. 
The values near zero momentum ${\bf k+G}=0$ are the
states with long wavelengths and reveal the macroscopic symmetry of
the crystal system, the 4-fold axis of the cubic system here.
If $|{\bf k+G}|$ is of the order of half
a reciprocal lattice vector, the interference with the next nearest neighbors
in the lattice becomes visible; the eight foothills in the figure
may be interpreted as a projection of the four corners of a carbon
tetrahedron onto the $(001)$ plane complemented by the inversion\cite{Falk}
\[
u_{\nu -{\bf k},-{\bf G}}= u_{\nu {\bf k},{\bf G}}^*.
\]
If $|{\bf k+G}|$ is large, the spherical symmetry of the core states prevails.

\subsection{Dielectric Function}
\begin{figure}[h]
 
 \includegraphics[scale=0.48]{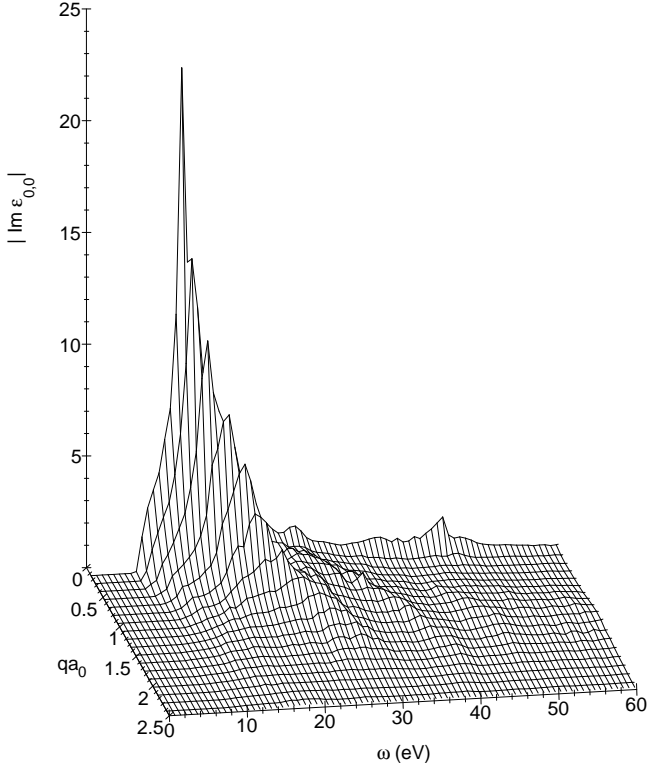}
\caption{
Calculated $\mathrm{Im} \epsilon_{\bf G,G'}$ for ${\bf G}={\bf G'}={\bf 0}$
and ${\bf q}\parallel (\bar 111)$.
The underlying DFT calculation is based on a $12\times 12\times 12$
mesh of ${\bf k}$-points
in the BZ (98 points in the IBZ), which creates for this particular
${\bf q}$-direction commensurate
values at $|{\bf q}|=(j/12)2\pi\sqrt{3}/a$ as shown here
for $j=1,2,3,\ldots$.
}
\label{fig3.ps}
\end{figure}

The element $\epsilon_{\bf 0,0}({\bf q},\omega)$ of the dielectric matrix
is shown in Fig.\ \ref{fig3.ps}\@.
The main absorption peak
at 11 eV for $|{\bf q}|\to 0$ corresponds to direct transitions from the top
of the valence to the bottom of the conduction band at X and L.\cite{Painter}

\begin{figure}[h]
 
 \includegraphics[scale=0.48]{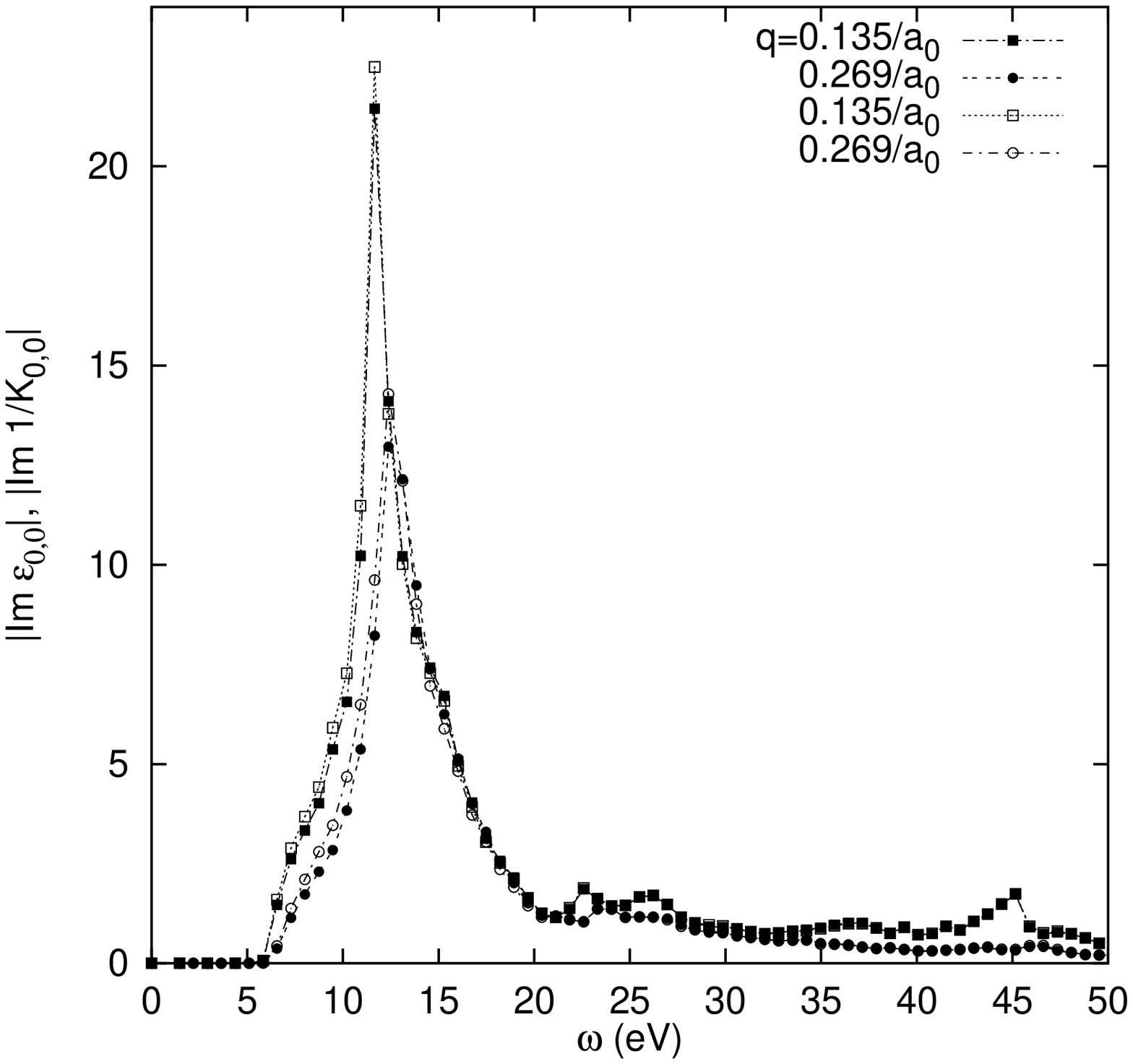}
\caption{
The absolute value of $\mathrm{Im}\epsilon_{\bf 0,0}({\bf q},\omega)$
(open symbols)
and of $\mathrm{Im}1/K_{\bf 0,0}({\bf q},\omega)$ (filled symbols).
${\bf q}$ is parallel to $(\bar 1 11)$ with $qa_0=0.135$ ($j=1$, squares) and
$qa_0=0.269$ ($j=2$, circles) as in Fig.\ \protect\ref{fig3.ps}.
Note that Refs.\ \protect\onlinecite{Vechten,GavriDia} and
\protect\onlinecite{HuangChing} refer to the optical limit $q=j=0$.
}
\label{fig4.ps}
\end{figure}

Local Field Effects are illustrated in Fig.\ \ref{fig4.ps}\@.
$\mathrm{Im} \epsilon_{\bf 0,0}$ repeats some values of Fig.\ \ref{fig3.ps};
$\mathrm{Im} [1/K_{\bf 0,0}]$ includes an estimate of the local field
by calculating $K_{\bf G,G'}$ as the inverse of a $9\times 9$ dielectric
matrix which contains ${\bf G}=0$ and the eight vectors
of the closest shell in the bcc reciprocal lattice.
The reduction of the values without LFE ($|\mathrm{Im} \epsilon_{\bf 0,0}|$,
open symbols) compared to those with LFE ($|\mathrm{Im} (1/K_{\bf 0,0})|$, filled
symbols) is of the order reported by Van Vechten and Martin\cite{Vechten}
(without their ``dynamical correlations'').
The different sign of the effect for frequencies above and below the peak has
been noticed before.\cite{GavriDia}
The differences are even smaller for the energy loss function.
Hence the energy loss in the next paragraph was
calculated from $\epsilon_{\bf 0,0}({\bf q},\omega)$ alone.

\subsection{Computed Electronic Energy Loss}

\begin{figure}[h]
 
 \includegraphics[scale=0.45]{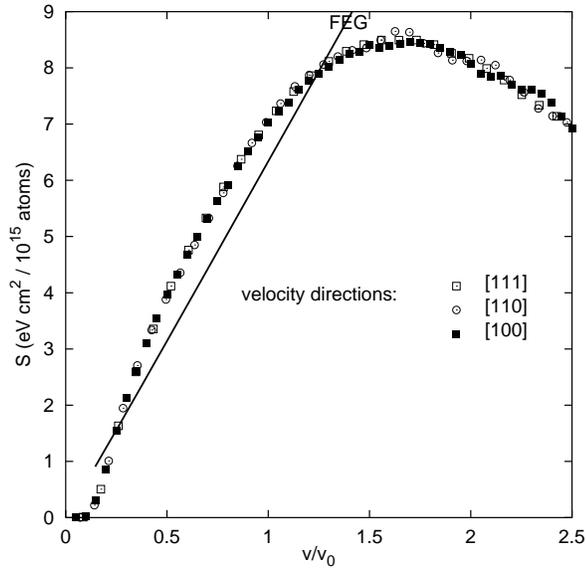}
\caption{
Electronic stopping cross section of diamond.
The stopping power was integrated on a mesh with 72 points on the
$\omega$-axis (0\ldots 103 eV). The lowest 28 bands were included
in the sum over band-pairs in Eq.\ (\ref{Piint}). The line refers to the
free electron gas \protect\cite{LindhardWi} with a density equivalent to
4 electrons per diamond atom.
}
\label{fig5.ps}
\end{figure}

\begin{figure}[h]
 
 \includegraphics[scale=0.45]{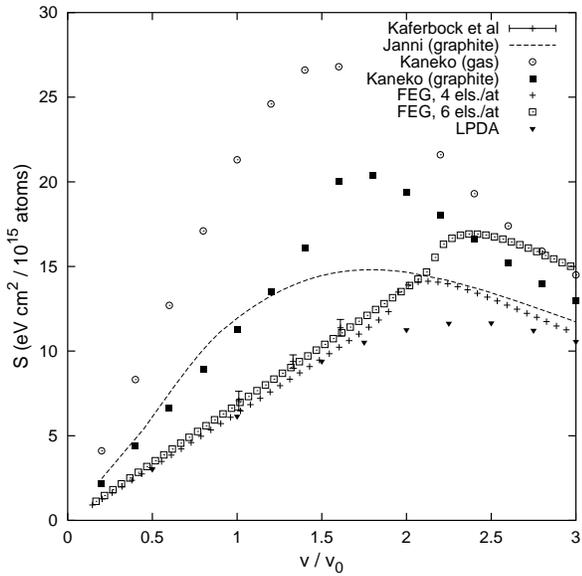}
\caption{
Experimental proton stopping cross section
by
K\"aferb\"ock {\it et al.\/} \protect\cite{Kaferb} and of graphite by Janni,%
\protect\cite{Janni} compared with calculated values for gaseous carbon and
solid graphite from Kaneko's theory,\protect\cite{Kaneko89} and RPA free
electron gas (FEG) values\protect\cite{LindhardWi} with a homogeneous density
equivalent to 4 or 6 electrons per diamond atom (Fermi velocity $1.457v_0$
or $1.668v_0$) as shown.
}
\label{fig6.ps}
\end{figure}

To integrate the energy loss function as described by Eq.\ (\ref{dedx}), the BZ
mesh was reduced to $8\times 8\times 8$ points, i.e., 35 points in the 
irreducible Brillouin zone (IBZ).
The result is shown in Fig.\ \ref{fig5.ps}\@.
The dielectric matrix was
tabulated on a $30\times 30\times 30$ mesh in ${\bf q}$-space
(parallelepiped with three edges of length $6.1/a_0$), and the stopping power
was integrated over the superset of all ${\bf q}$-values obtained
from these via point group operations with Seitz symbol $\{O|{\bf w}\}$,
\[
K_{\bf G,G'}({\bf q},\omega)=
e^{iO({\bf G-G'})\cdot {\bf w}}K_{O{\bf G},O{\bf G'}}(O{\bf q},\omega) .
\]
It is about 0--15\% lower than experimental results\cite{Kaferb} for
$v\approx 1\ldots 1.3v_0$ which are shown in Fig.\ \ref{fig6.ps}\@.
($v_0\approx 2.19\cdot 10^6$ m/s is one atomic unit of velocity.
Values within the LPDA represent intgrals of the all-electron density of
{\sc gtoff} Fitting Functions weighted with the stopping number of the
FEG in the RPA\@.)
Reduction of the integrated ${\bf q}$-region to the parallelepiped
decreases the stopping cross section by approximately 10\% at
$v\approx v_0$.\cite{wbw18}
Within the framework of linear dielectric response an underestimation is
appropriate though,
because terms of $O(Z_1^3)$ will add about 15\%
to $S(v)$ at $v\approx 1.5v_0$.\cite{Jackson}

A new result is the ``ionic'' band gap, the
non-linear suppression of $S(v)$ for $v\lesssim 0.2v_0$,
which is approximately the value extracted from\cite{wbw18}
\begin{equation}
v < \frac{v_0}{2} \sqrt{\hbar \omega_g/E_0}
\label{vgap}
\end{equation}
for a band gap of $\hbar\omega_g=4$ eV\@.
($E_0$ is one rydberg.)
By way of contrast, calculations within the
local-plasma-density approximation usually integrate over volume elements in
real space that are parameterized by the homogeneous electron gas,
and inevitably yield $S(v)\propto v$ at low velocities.
However, an attempt at experimental verification of this reduction of the
energy loss in a wide-gap material is handicapped by the additional
nuclear energy loss, which has an estimated maximum of
$S\approx 0.8\cdot 10^{-15}$ eVcm$^2/$atoms at
$v\approx 0.08v_0$.\cite{Littmark}

Note that Eq.\ (\ref{vgap}) predicts
contributions from $K$-shell excitations to start at $v\approx 2.2v_0$,
which are not included here. They have been estimated to shift
the maximum to higher velocities by about $0.2v_0$,\cite{SabinDia}
and to grow until $v\approx 5.3v_0$.\cite{Paul}

\begin{acknowledgments}
 Helpful conversations with S.B. Trickey and J.R. Sabin
 are gratefully acknowledged.
 This work was supported by grant DAA-H04-95-1-0326 from
 the U.S.\ Army Research Office.
\end{acknowledgments}

\appendix

\end{document}